\documentclass[a4paper,fleqn,usenatbib]{mnras}

\usepackage[T1]{fontenc}
\usepackage{ae,aecompl}

\usepackage{graphicx}	
\usepackage{amsmath}	
\usepackage{amssymb}	

\usepackage{txfonts}

\title[Early-types galaxies and the MDA relation]{The mass discrepancy acceleration relation in early-type galaxies: extended mass profiles and the phantom menace to MOND}
\author[J. Janz et al.]{
Joachim Janz,$^{1}$\thanks{E-mail: jjanz@swin.edu.au}
Michele Cappellari,$^{2}$
Aaron J. Romanowsky,$^{3,4}$
Luca Ciotti,$^{5}$\newauthor
Adebusola Alabi,$^{1}$
and Duncan A. Forbes$^{1}$
\\
$^{1}$Centre for Astrophysics \& Supercomputing, Swinburne University, Hawthorn, VIC 3122, Australia\\
$^{2}$Sub-department of Astrophysics, Department of Physics, University of Oxford, Denys Wilkinson Building, Keble Road, Oxford OX1 3RH, UK\\
$^{3}$Department of Physics and Astronomy, San Jos\'e State University, San Jose, CA 95192, USA\\
$^{4}$University of California Observatories, 1156 High Street, Santa Cruz, CA 95064, USA\\
$^{5}$Department of Physics and Astronomy, University of Bologna, I-40127 Bologna, Italy
}

\date{Accepted 2016 June 15. Received 2016 June 15; in original form 2016 March 6}

\pubyear{2016}

\begin{document}
\label{firstpage}
\pagerange{\pageref{firstpage}--\pageref{lastpage}}
\maketitle

\begin{abstract}
The dark matter (DM) haloes around spiral galaxies appear to conspire with their baryonic content:
empirically, significant amounts of DM are inferred only below a universal characteristic acceleration scale.
Moreover, the discrepancy between the baryonic  and dynamical mass, which is usually interpreted
as the presence of DM, follows a very tight mass discrepancy acceleration (MDA) relation. 
Its universality, and its tightness  in spiral galaxies, poses a challenge for the DM interpretation 
and was  used to argue in favour of MOdified Newtonian Dynamics (MOND).
Here, we test whether or not this applies to early-type galaxies.  We use the dynamical models 
 of fast-rotator early-type galaxies by Cappellari et al.\ based on ATLAS$^{\textrm{3D}}$ and SLUGGS data, 
which was the first homogenous study of this kind, reaching  $\sim$$4R_\textrm{e}$, where DM begins to dominate the total mass budget. 
 We find the early-type galaxies to follow an  MDA relation similar to spiral galaxies, 
 { but systematically offset. Also, while the slopes of the 
 mass density profiles inferred from galaxy dynamics 
 show consistency with those expected from their stellar content assuming MOND, some profiles of
individual galaxies show discrepancies. }

\end{abstract}




\section{Introduction}
The masses of galaxies, and in fact of any larger dynamically bound structure in the Universe, 
inferred from dynamics are found to exceed  the masses of the observed baryons in
these structures
\citep{1933AcHPh...6..110Z,1937ApJ....86..217Z,1970ApJ...159..379R,Rubin:1980dm,Bosma:1978vi}.
This phenomenon is usually explained by postulating
(non-baryonic) dark matter, which is also the backbone of structure formation 
in cosmological simulations \citep[e.g.][]{1985ApJ...292..371D,2009MNRAS.398.1150B}
and serves well in accounting for the characteristics of the 
cosmic microwave background radiation
\citep[e.g.][]{1984ApJ...285L..45B,2015arXiv150201589P}. 

~\\

\citet{1990A&ARv...2....1S} demonstrated a surprising characteristic of this discrepancy between dynamical 
and baryonic mass: it  occurs below a characteristic acceleration level.
\citet{2000ApJ...534..146V} reproduced such a characteristic acceleration scale in 
the $\Lambda$CDM context using semi-analytic models of galaxy formation, which were
tuned to reproduce the Tully-Fisher relation. 
However, \citet{1990A&ARv...2....1S} also found  that the amplitude of this discrepancy  
correlates with the acceleration, the so-called mass discrepancy acceleration (MDA) relation. 
Moreover, there is only small scatter about this relation, 
as confirmed by \citet{2004ApJ...609..652M}.
This  correlation, and even more its tightness, is surprising in the context of hierarchical
structure formation and the large variety of possible merger trees
for individual galaxies in $\Lambda$CDM cosmology (see also \citealt{2004ApJ...609..652M,2014Galax...2..601M,2014ConPh..55..198W}).

~\\

So far tests of the MDA relation have concentrated mainly on spiral galaxies. In these systems the H\,\textsc{i} 
gas makes it relatively easy to trace the dynamical mass far  from the centre of the galaxy.
However, spiral galaxies also possess large amounts of baryonic mass in the form of gas. The mass of this component is rather difficult to estimate accurately from observations and introduces uncertainties in the MDA relation.
Early-type galaxies  (ellipticals and lenticulars), due to their lower H\,\textsc{i} gas content, do not suffer from this problem, however they lack an easy to measure tracer. For this reason, studies addressing the MDA relation in early-type galaxies have had to largely rely on other tracers
such as hot X-ray emitting gas or discrete tracers like planetary nebulae (PNe), globular clusters (GC), 
and 
satellite galaxies (e.g.\ \citealt{2012PhRvL.109m1101M}; { see however \citealt{2001AJ....121.1936G}}). 
While theoretical efforts trying to reproduce the MDA relation in the $\Lambda$CDM context are progressing 
(e.g.~\citealt{2016MNRAS.456L.127D,}), it is desirable to push
the observational side further, and probe the MDA relation systematically for a sample of early-type galaxies { down to low accelerations.}

~\\

Alternatively, the MDA relation, and its tightness, have been used to argue in favour of a modification of  Newtonian dynamics (MOND; see, e.g.~\citealt{2012LRR....15...10F}; \citealt{2015MNRAS.446..330W}).
A number of previous studies  have discussed whether early-type galaxies can be used to falsify the theory
\citep{2003ApJ...599L..25M,2007A&A...476L...1T,2008MNRAS.383.1343W,2009ApJ...690.1488K,2011A&A...531A.100R,2012LRR....15...10F,2012A&A...538A..87S,2014A&A...570A.132S,2015MNRAS.451.1719C,2015A&A...581A..98D}.

\defcitealias{Cappellari:2015gt}{C+15}
Recently, \citet[hereafter \citetalias{Cappellari:2015gt}]{Cappellari:2015gt} carried  out dynamical modelling of a sample of 14
early-type galaxies, empowered by the combination of the inner stellar kinematics from ATLAS$^{\textrm{3D}}$ \citep{2011MNRAS.413..813C} and stellar kinematics reaching out to 
a median radius of about 4 half-light radii  ($R_\textrm{e}$; \citealt{2014ApJ...791...80A}) from the SLUGGS survey  \citep{SLUGGS}.
This represents the first homogeneous, 
 statistically meaningful sample of stellar kinematics of early-type galaxies (all of which are fast rotators as defined in \citealt{Emsellem:2011br}) reaching radii where dark matter is expected to dominate the mass budget { (i.e. the dark matter fraction increases from $\sim$1/3 to $\sim$2/3 when measured within 2  and 4 $R_{\rm e}$, respectively).} 

Here we use \citetalias{Cappellari:2015gt}'s mass modelling results 
to test whether or not early-type galaxies follow the MDA relation. Furthermore,
we test whether or not MOND, which was originally designed to reproduce the rotation curves of spiral galaxies \citep{1983ApJ...270..365M}, fails in early-type galaxies.
{ Finally, we also consider results from mass modelling based on GC kinematics for an extended sample, including slow rotators and reaching even larger radii, also from the SLUGGS survey \citep{alabi}.}

\section{Sample, Data and Dynamical Modelling}

Our primary sample comprises 14 early-type galaxies in a stellar
mass range of $1.5 \times 10^{10} < M_* / \textrm{M}_\odot < 50 \times 10^{10}$ \citep{2013MNRAS.432.1709C}.
Mass models for these galaxies were built by \citetalias{Cappellari:2015gt}
using axisymmetric Jeans anisotropic modelling (JAM; \citealt{2008MNRAS.390...71C}) with a stellar and a dark matter component. To ensure the method to be
applicable, and to achieve a homogenous sample of nearly axisymmetric galaxies, \citetalias{Cappellari:2015gt} 
restricted the sample to fast rotators. 

One input to the JAM modelling is the light distributions in these galaxies. Those were 
parametrized using Multi-Gaussian Expansions (MGE; \citealt{1994A&A...285..723E,2002MNRAS.333..400C})
and were taken from various studies 
(\citealt{1999MNRAS.303..495E,2006MNRAS.366.1126C,2009MNRAS.398.1835S,2013MNRAS.432.1894S}; \citetalias{Cappellari:2015gt}).
Here, we use the same characterizations of the distributions of stars within the galaxies.
Other galaxy parameters, such as half-light radius  and distance, are taken from \citetalias{Cappellari:2015gt}, and the velocity dispersion within $\sigma_{\rm e}$ from \citet{2013MNRAS.432.1709C} and \citet{2013ARA&A..51..511K} for NGC~3115.

The combination of data from ATLAS$^{\textrm{3D}}$ \citep{2011MNRAS.413..813C} and SLUGGS  \citep{SLUGGS} allowed \citetalias{Cappellari:2015gt} to probe the stellar dynamics of the galaxies from their inner parts out to a median radius of $\sim$4\,$R_{\textrm{e}}$. The stellar kinematics were extracted in the optical (ATLAS$^{\textrm{3D}}$; \citealt{2004MNRAS.352..721E,2011MNRAS.413..813C}) and Ca triplet spectral region (SLUGGS; \citealt{2014ApJ...791...80A}), in both cases
with pPXF \citep{2004PASP..116..138C}. For the dynamical modelling the data were symmetrized, outliers were removed, and the two data sets  combined. We refer the reader to 
\citetalias{Cappellari:2015gt} for a description of the process. However, we do note the key characteristic of \citetalias{Cappellari:2015gt} was the use of a very general parametrization for the dark halo in the modelling process, to be able to focus on the {\em total} density profile alone.

\begin{figure}
	\includegraphics[height=0.95\columnwidth,angle=-90]{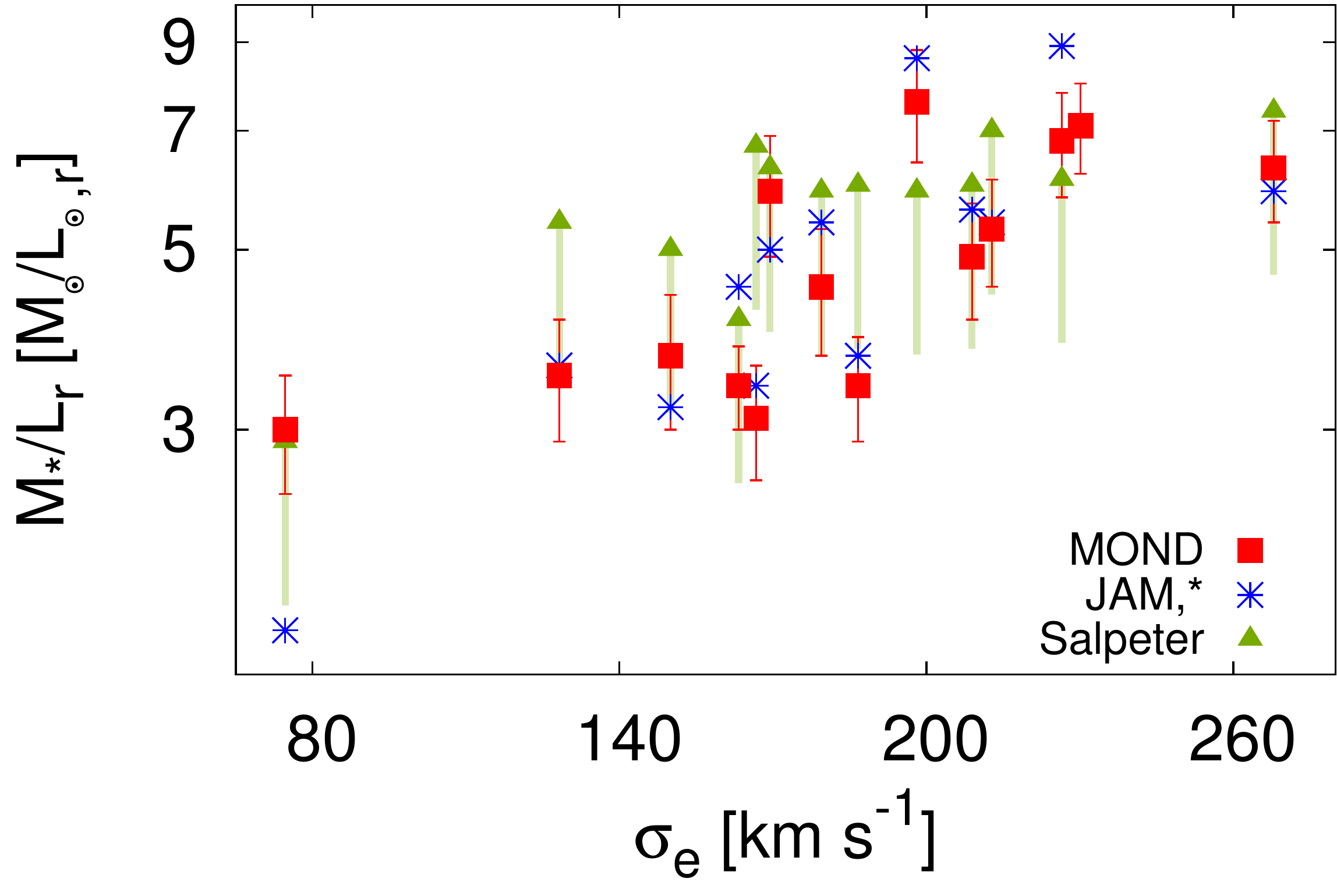}
     \caption{Comparison of stellar  mass-to-light ratios. The red squares  show the mass-to-light ratios from 
     fits of the dynamical models to the observations assuming MOND ( $(M/L)_{\rm MOND}$; see also Section \ref{sec:MOND}).
      Blue stars indicate the stellar $(M/L)_{{\rm JAM},*}$ obtained directly  from  JAM of ATLAS$^\textrm{3D}$ data \citep{2013MNRAS.432.1862C}.   The stellar population mass-to-light ratios from the same study, which assume a Salpeter IMF, are shown as green triangles, and the lines indicate a factor of 0.63 
     approximating the corresponding value for a Kroupa IMF.
     The various methods for calculating stellar mass-to-light ratios reproduce the same qualitative trend of stellar mass-to-light ratio with velocity dispersion $\sigma_e$.}
    \label{fig:ML}
\end{figure}

The JAM models are very accurate and robust, which was confirmed by comparison with the results of \citet{Serra16} based on H\,\textsc{i}  data, but despite the increase in radial extent when compared to similar earlier studies, they probe the dynamics to accelerations an order of magnitude higher than those typically probed in spiral galaxies.
\citet{alabi} recently applied the  tracer mass estimator of \citet{2010MNRAS.406..264W} to the kinematics of the GC systems for a superset of early-type galaxies, mostly from the SLUGGS survey \citep{SLUGGS}. 
This method estimates the enclosed mass from the line-of-sight velocities of the tracer population based on  assumptions for the power-law slope of the gravitational potential, the power-law slope of the tracer density profile, and the orbital anisotropy. 
While being less reliable, these models allow us to get indications about trends   beyond  5\,$R_{\textrm{e}}$, and also for slow rotators.
\citeauthor{alabi} analysed the dependence of their mass estimates on these assumptions, took into account corrections for non-sphericity and substructures in the tracer population, and compared their results to previous studies. 
Their total mass estimates with the assumption of isotropy  agree  with similar earlier studies within a factor of 1.6. 
We refer the reader for the details to \citeauthor{alabi} and adopt their estimates for isotropic orbits. 

The SLUGGS galaxies were selected to be representative early-type galaxies across various environments (see  \citealt{SLUGGS}) and
span a wide mass range  up to NGC~4486, { i.e.~M87.}
For 23 of the 25 SLUGGS galaxies \citeauthor{alabi} derived estimates for the total mass within $5\,R_{\textrm e}$ and beyond.
Most of the galaxies have rich enough GC data sets to derive not only  a single dynamical mass estimate, but also radial mass density profiles. This applies for 10 of the 14 galaxies of our primary sample.  Beyond the galaxies common to both samples, the study of \citeauthor{alabi}  comprises NGC~1400, NGC~3607, NGC~4564, NGC~5866 and also the slow rotators NGC~720, NGC~1407, NGC~3608, NGC~4365, NGC~4374, NGC~4486, and NGC~5846.

\section{MDA relation in early-type galaxies}

Before analysing the MDA relation, we need to convert the stellar light to stellar mass,
in order to estimate the expected accelerations.
Several options are available. The stellar populations can be fitted with models to infer
a stellar mass-to-light ratio. This requires knowledge of the stellar populations and needs to assume
a stellar initial mass function (IMF), which is suspected to vary from galaxy to galaxy \citep[e.g.][]{2010Natur.468..940V,Cappellari:2012jm}, even in the framework of MOND \citep{2014MNRAS.438L..46T}.
The stellar mass-to-light ratio $(M/L)_{{\rm JAM},*}$ can also be obtained from the dynamical model. For ATLAS$^\textrm{3D}$, the ratio is then based on the assumption of Newtonian gravity \citep[denoted as $(M/L)_{\rm stars}$ therein]{2013MNRAS.432.1862C}.
Here, we also compute the stellar mass-to-light ratio  $(M/L)_{\rm MOND}$  by fitting the  dynamics expected
from the light profile based on MOND to  the observed dynamical profile, and with a spatially constant mass-to-light ratio as fitting parameter.
This fit is dominated by the inner regions where density and flux are highest.

We compare the various mass-to-light ratios in Fig.~\ref{fig:ML}.
Since the MGE used in the dynamical modelling came from observations with different
photometric filters, we convert all mass-to-light ratios to the $r$-band, using the photometric predictions of \citet{2012MNRAS.424..157V} and \citet{Ricciardelli:2012cw}. 
The various ways of determining the mass-to-light ratio all show the same trend of an increase with
increasing velocity dispersion $\sigma_{\rm e}$ within $R_{\rm e}$ \citep[see also, e.g.,][]{2006MNRAS.366.1126C,2007ApJ...668..756V}.
This trend is due to underlying changes of the
stellar population characteristics, e.g.~increasing age and metallicity (and increasing $M/L$ due to changes in the IMF).
In the following we use the mass-to-light ratios from the MOND fitting $(M/L)_{\rm MOND}$ for consistency with the comparisons in Section \ref{sec:MOND}.

In the next step the enclosed mass is calculated, both for the dynamical mass from  JAM,
which is calculated from the density profiles of \citetalias{Cappellari:2015gt}, and for the stellar
mass. 
The latter is obtained from the published MGEs of the galaxy light distribution and the spatially constant stellar mass-to-light ratio $(M/L)_{\rm MOND}$ 
 as follows.
The mass of an axisymmetric MGE model, enclosed within a spherical shell of radius $r$ is given by
\begin{equation} 
\begin{split}
& M(r) =  \sum_{i} M_{i}  \left  \{ \textrm{erf}\left[r/(\sqrt{2}q_i\sigma_i)\right] \,\,\, - \vphantom{ { { \exp\left[-r^2/(2\sigma_i^2)\right]  \textrm{erf}\left[r \sqrt{1-q^2_i}/(\sqrt{2}q_i\sigma_i) \right] } \bigg/{  \sqrt{1-q^2_i} }}} \right. \\
 & \left. { { \exp\left[-r^2/(2\sigma_i^2)\right]  \textrm{erf}\left[r \sqrt{1-q^2_i}/(\sqrt{2}q_i\sigma_i) \right] }\bigg/ {  \sqrt{1-q^2_i} }} \right \} ,
\end{split}
\end{equation}
with the MGE parameters for width, flattening, and total mass  ($\sigma_i$, $q_i$, and $M_{i}$), for each Gaussian component respectively.
This equation was obtained by integration of the density profile in terms of the MGE as given in footnote 11 
of \citetalias{Cappellari:2015gt}.

In addition to stellar and dynamical mass,  the Newtonian acceleration due to the baryons is needed for the MDA relation, and it is given in the spherical limit by
\begin{equation} 
g_{N,*}(r) = G M_*(r) / r^2.
\label{eq:Newton}
\end{equation}

We calculate uncertainties 
in a Monte Carlo fashion, by randomly perturbing the density profiles of 
 \citetalias{Cappellari:2015gt}
within their errors (see Sec.~\ref{sec:MOND}). 
However, rather than using only formal uncertainties, we try to be conservative and account for systematic uncertainties as follows.
We adopt a 6\% uncertainty in the overall mass normalization as inferred by \citet{2013MNRAS.432.1709C}
and an upper limit of 0.11 in the profile slope, derived from the observed scatter in  \citetalias{Cappellari:2015gt}.
Therefore, we perturb the profiles by adding a random constant with $\sigma = 0.025$ dex to log$(r)$, and by adding a linear trend in $\log\rho(r)$ versus $\log(r)$  with a random slope ($\sigma=0.05$). We calculate and plot for each galaxy 100 realizations of this process.

\begin{figure*}
	\includegraphics[width=0.95\columnwidth,angle=0]{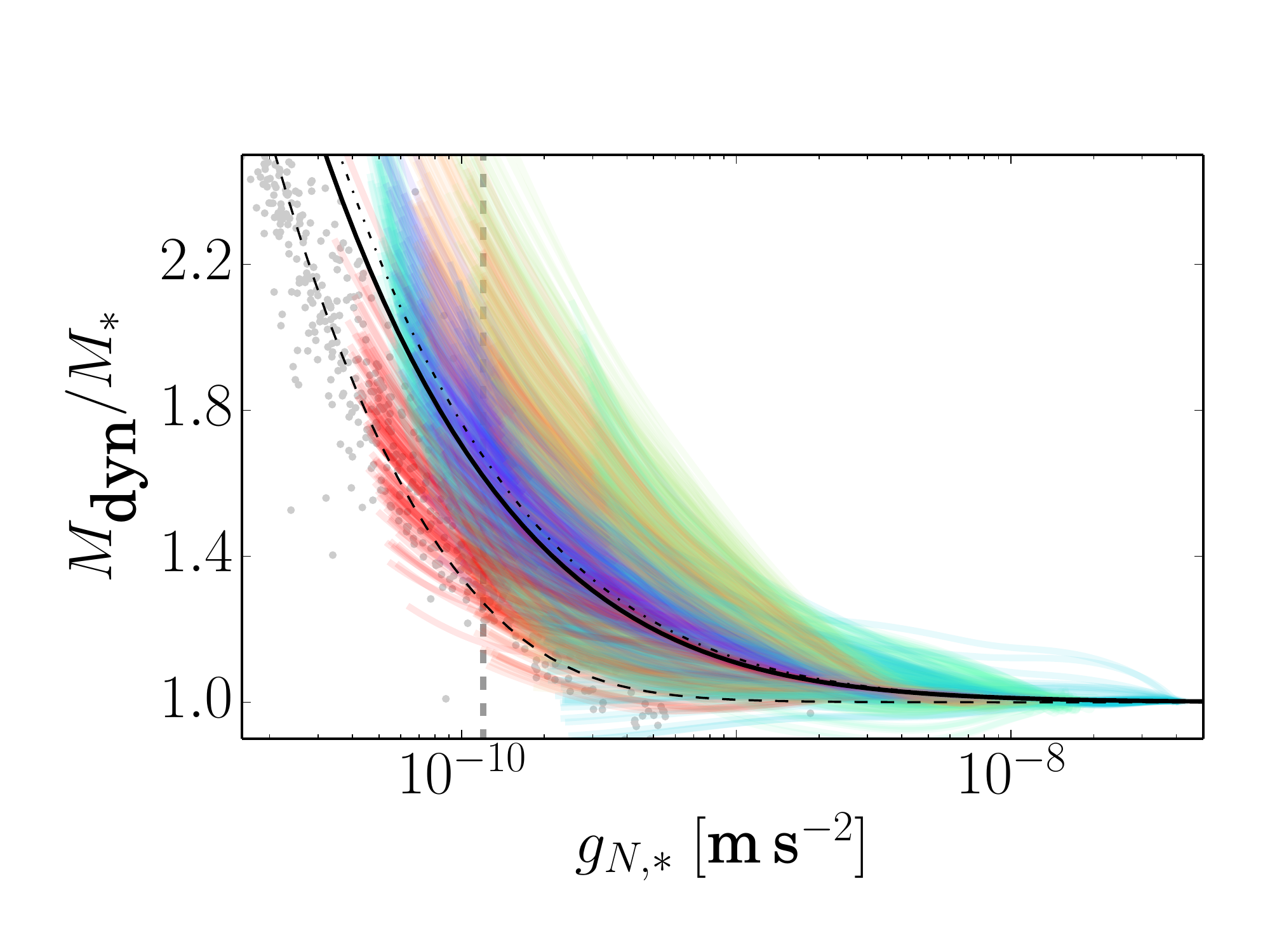}
	\includegraphics[width=0.95\columnwidth,angle=0]{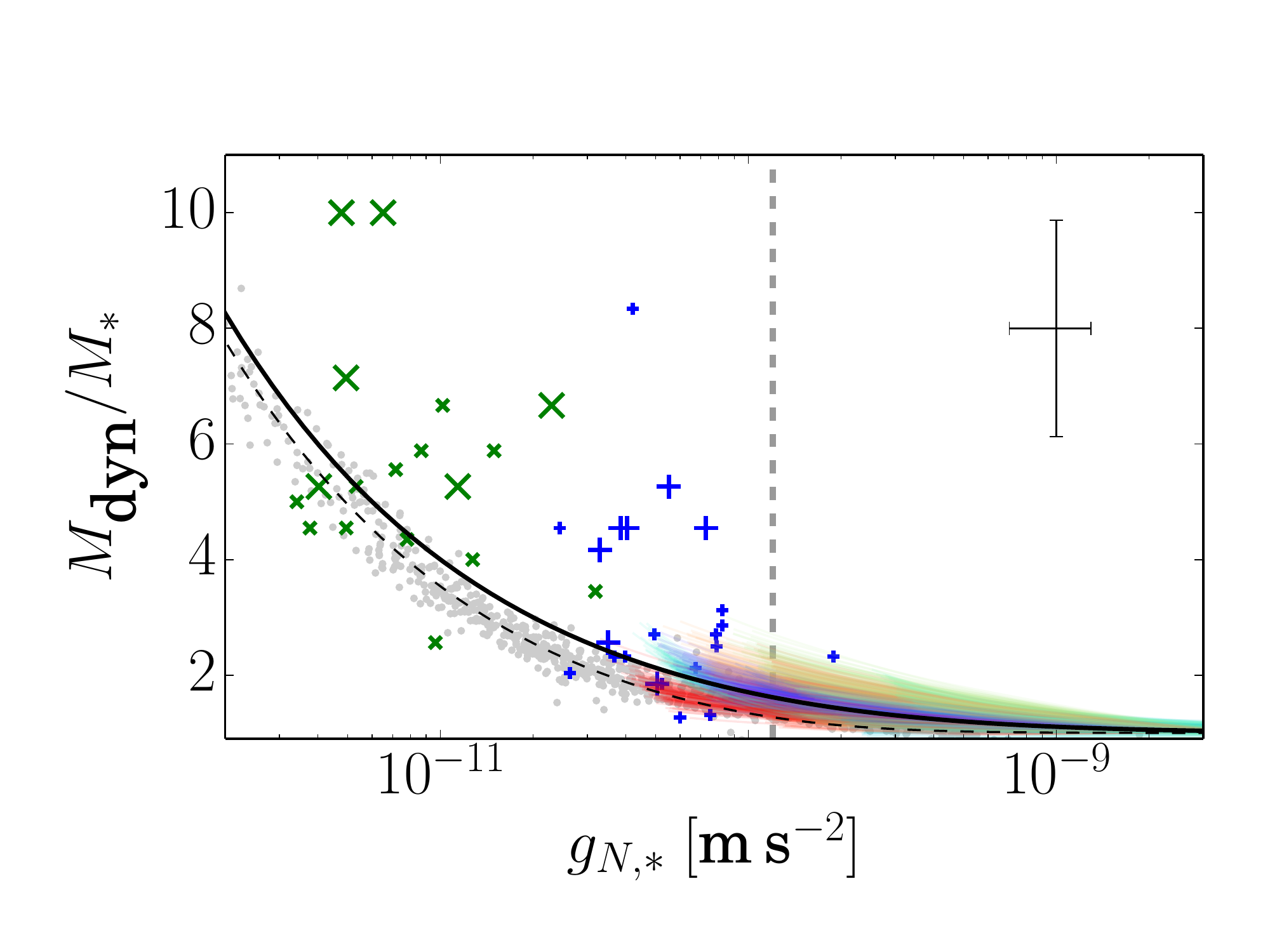}
     \caption{Mass discrepancy acceleration relation. {\it Left} panel: the dynamical-to-stellar mass ratio  versus the acceleration caused by the stars (both quantities inferred through Newtonian dynamics). {\it Right} panel: MDA relation as in the other panel. Results based on GC system dynamics from \citet{alabi} are shown as blue plus signs and and green crosses for values at $5 R_\textrm{e}$ and $R_\textrm{max}$, respectively (see text), and larger symbols mark slow rotators. Typical error bars are indicated in the top right corner (these show random variations, while additional systematics of similar magnitude can be expected). For comparison, data for spiral galaxies from \citet{2012LRR....15...10F} are included in both panels as grey points. The { different colours of the lines denote} the various galaxies,  each with 100 realizations  (see text). { The black { solid} curve shows the relation as expected from MOND using the {\it simple} interpolating function (see Equation \ref{eq:massr}), and the grey vertical dashed line indicates the corresponding value of $a_0 = 1.2 \times 10^{-10}$ m s$^{-2}$.  The dash-dotted curve { in the left panel} shows the same MOND relation for $a_0 = 1.35 \times 10^{-10}$ m s$^{-2}$, and the dashed curve that for   the {\it standard} interpolating function  with  $a_0 = 1.2 \times 10^{-10}$ m s$^{-2}$.} { While our sample follows an MDA relation with large scatter, it is offset from the comparison sample of spiral galaxies. The slow rotators in the right panel display still larger offsets. }  \label{fig:MDA}}
\end{figure*}

The resulting MDA relation is plotted for all 14 galaxies in the left-hand panel of Fig.~\ref{fig:MDA}. It shows that our fast rotators follow a relation similar to that found for spiral galaxies, with the dynamical to stellar mass discrepancy systematically increasing with decreasing acceleration at that radius.

\section{Comparison to modified Newtonian Dynamics}
\label{sec:MOND}
\begin{figure*}
	\includegraphics[width=\columnwidth,angle=0]{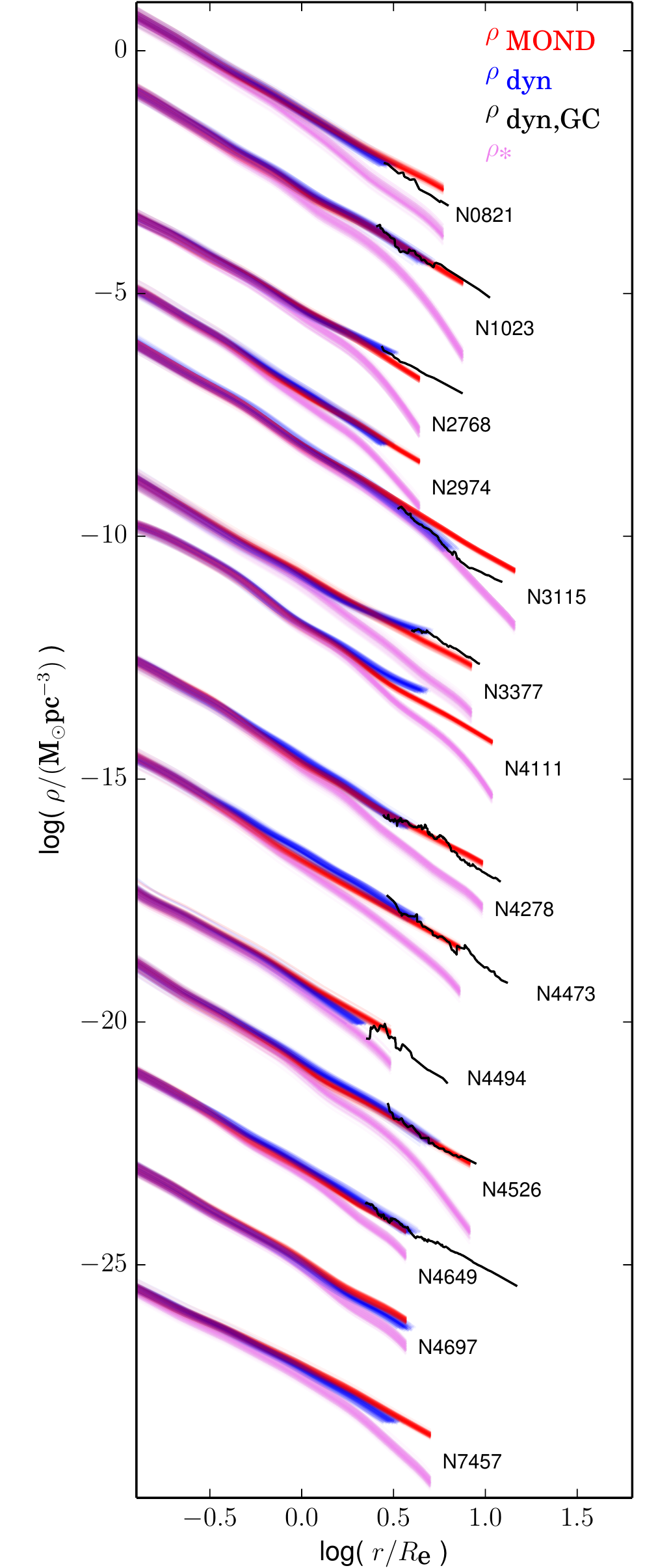}
	\includegraphics[width=\columnwidth,angle=0]{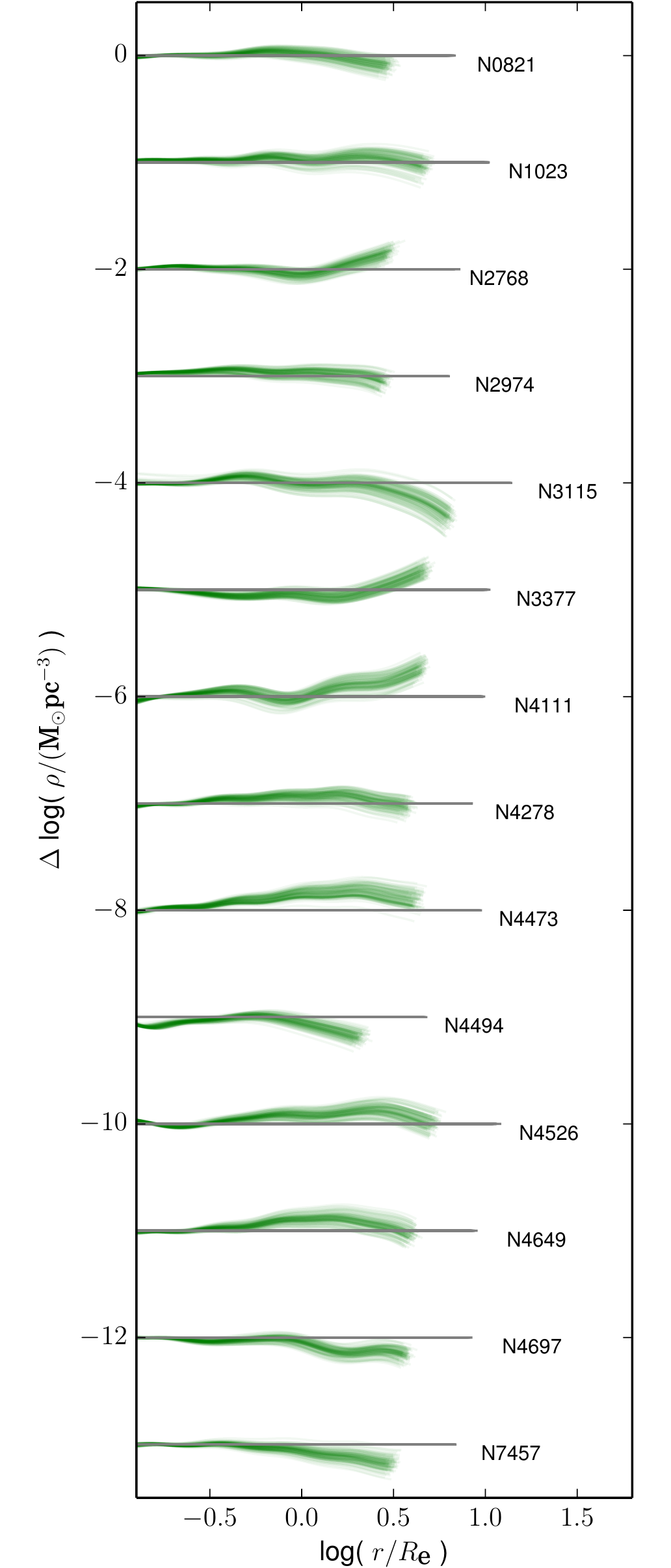}
    \caption{Mass density profiles. {\it Left: } Shown  for each galaxy are 100 representations of the density profiles obtained as described in the text. The various colours show: the dynamically determined density profile from  \citet[blue]{Cappellari:2015gt}, the stellar density profile (lavender) from the published MGE using a spatially constant mass-to-light ratio from fitting the MOND predictions (red) to the dynamically obtained density profiles. In addition,  where available, we plot the density profile determined using tracer mass estimators of the GC systems (black) with vertical shifts to match the densities in the overlap region (see text). Vertical offsets are applied to the profiles by 2 units per galaxy. { \textit{Right:} The difference between the dynamically determined and MOND predicted mass profiles is shown, with vertical offsets applied to the profiles by 1 unit per galaxy. { The green curves show the Monte Carlo realizations for this difference, while the grey lines indicate the zero difference level for each galaxy.} { While the MOND predictions generally produce near-isothermal density profiles as observed, the consistency between MOND predictions and the dynamically determined density profiles is less convincing for individual galaxies.}}\label{fig:profiles}} %
\end{figure*}

\citet{1983ApJ...270..365M} introduced
a modification to Newtonian dynamics  
as an alternative to dark matter  in explaining   flat rotation curves in spiral galaxies
(for  recent reviews see \citealt{2012LRR....15...10F}; \citealt{2015CaJPh..93..119B}, and the entire Special Issue in which it was published).

In Newtonian dynamics the acceleration caused by a spherical matter distribution  within a radius $r$ is given by Equation (\ref{eq:Newton}).
In  MOND the acceleration  $a$
felt by matter is modified. Under the assumption of spherical symmetry the two quantities are related by
\begin{equation}
g_N = \mu(a/a_0)\ a,
\label{eq:mu}
\end{equation}
with the interpolating function $\mu$ and a characteristic acceleration scale
$a_0 = 1.2 \times 10^{-10}$ m s$^{-2}$. For high accelerations ($a\gg a_0$)
$\mu \rightarrow 1$, i.e.~the modification vanishes and the Newtonian limit
applies.
Equation (\ref{eq:mu}) can be easily inverted
\begin{equation}
a=\nu\left(g_N/a_0\right)g_N,
\label{eq:acc}
\end{equation}
with $\mu(x) = 1/\nu[x\mu(x)]$. 
One common choice for the interpolating function $\mu(x)$, especially on galactic scales,
is the \textit{simple}  interpolating function 
$\mu(x)=x/(1+x)$, so that
 $\nu(y)=1/2 + \sqrt{1/4 + 1/y}$.

 The dynamically inferred mass $M_\textrm{dyn}$, when assuming Newtonian dynamics, is then 
 given by
 \citep[see, e.g.,][]{2012PhRvL.109m1101M}

\begin{equation}
M_\textrm{dyn}(r) = a(r)\ G^{-1} r^2  = \nu\left[ \frac{G M_*(r)}{r^2 a_0}  \right] M_*(r).
\label{eq:massr}
\end{equation}
This offers a natural explanation for flat rotation curves in disc galaxies and for the baryonic Tully-Fisher relation (see, e.g., \citealt{2012LRR....15...10F}), 
and it means a Newtonian observer concludes the presence of -- from the MOND point of view `phantom' -- dark matter given by $M_\textrm{DM} = M_\textrm{dyn} - M_*$.

In terms of the MDA  this leads to a simple prediction, which is plotted as the black { solid} curve in Fig.~\ref{fig:MDA} (see also, e.g.,  \citealt{2015CaJPh..93..169K} for a more theoretical account). 
{ Given the uncertainties the MOND curve looks like a rather good representation for our sample.

\citetalias{Cappellari:2015gt} found the galaxies to have nearly isothermal mass density profiles ($\rho(r) \sim r^{-2.2}$), with only small amounts of scatter, which was  confirmed by  \citet{Serra16} for a sample of early-type galaxies based on H\,\textsc{i} data.
In Fig.~\ref{fig:profiles} the stellar density profile, the dynamical mass density profile  from  \citetalias{Cappellari:2015gt}, and that predicted from the stellar density profile using MOND are shown.
 It can be seen that the MOND profiles reproduce the slopes of the dynamically determined profiles well.
This is not a surprise, since galaxies have asymptotically isothermal profiles in  MOND.
 However, when  compared in more detail, for each galaxy individually,  inconsistencies become noticeable.
There are galaxies like NGC~3115 that start to diverge strongly from the MOND prediction  at the outermost radii,
but also galaxies like NGC~4473 and NGC~4494 for which features at smaller radii are not well reproduced. This becomes especially evident
when considering the residual profiles in the right panel, which are calculated individually for each Monte Carlo representation.}

\section{Discussion} 
The general reproduction of the MDA and the slopes of the density profile by MOND is intriguing at first.  
We note that this is achieved without fitting for the MOND acceleration scale, but using  for all galaxies a value of $a_0=1.2\times10^{-10}$ m s$^{-2}$, which was found
by \citet{1991MNRAS.249..523B} for spiral galaxies, and which is still used in more recent studies \citep[e.g.][]{2012LRR....15...10F}.
The sensitivity to the exact value is not very critical, as can be seen in the comparison curve in the MDA relation for another value $a_0=1.35\times10^{-10}$ m s$^{-2}$, which was also used for spiral galaxies (e.g.~\citealt{Famaey07}).

However, as \citetalias{Cappellari:2015gt} showed, the dynamical profiles can also be expected
to be simple power-laws in the $\Lambda$CDM context out to about 10 $R_{\textrm{e}}$.
Furthermore, the profile for individual cases are less consistent as mentioned above.

Our analysis makes a few potentially limiting assumptions, such as a spatially constant mass-to-light ratio and  anisotropy. \citetalias{Cappellari:2015gt} accounted for two individual anisotropies in the inner and outer part.
Also, for the calculation of the MONDian predictions we assumed spherical symmetry, which may seem to be a rather strong assumption given that the galaxies in our sample are -- often flattened -- fast rotators. 
\citet{Ciotti:2006jf} and \citet{2007MNRAS.379..597N} demonstrated that the expected deviations due to non-sphericity effects on the non-linear MOND equivalent of Poisson equation are however small even in flat spiral galaxies. 
The modelling uncertainties are dominated by systematics, which can be important in individual cases. For this reason significant progress can best be made in a statistical manner using large galaxy samples.

With this in mind, the following could be considered as a  more serious challenge for MOND than the mismatches for individual galaxies.
We showed our comparisons with the MOND predictions when using the simple interpolating function, since it gives  more consistent results for our sample than the \textit{standard} one  \citep[cf. also][]{2005MNRAS.363..603F,2007MNRAS.379..702S,2008MNRAS.383.1343W,2012PhRvL.109m1101M,2015MNRAS.451.1719C}. While changing the MOND constant $a_0$ within the range of values used for spiral galaxies only leads to minor changes, switching to the standard interpolating function has more severe effects.
The number of density profiles that are still consistent with MOND given our assumptions and estimates of uncertainties is roughly halved as compared to that when using the simple interpolating function.
In the MDA relation this is even more evident. The calculation with the standard interpolating function largely  underpredicts the mass discrepancy for our sample (Fig.~\ref{fig:MDA}) and marks essentially the lower edge of the trend of observed  MDA curves, while the MDA relation calculated with the simple interpolating function runs through the middle of this trend. However, the opposite is true for 
the comparison sample 
of spirals from \citet{2012LRR....15...10F}, for which the standard interpolating function provides a superior representation. This is at odds with MOND, where there should be one universal interpolating function and MDA relation. 

{ The above is consistent with
\citeauthor{2001AJ....121.1936G}'s (\citeyear{2001AJ....121.1936G}) results from stellar dynamics.  They 
concluded that, from their analysis of the MDA  for a sample of early-type galaxies, 
 the upturn of the MDA} relation occurs at higher accelerations than for spiral galaxies, in disagreement with MOND
(when using accelerations $a=v^2/r$ based on the dynamics as in their fig.~19 our Fig.~\ref{fig:MDA} looks similar). 
Their sample included two of our  galaxies  and  also slow rotators. At that time, 
\citet{2003ApJ...599L..25M} doubted that the change in the mass-to-light ratios, which is what \citeauthor{2001AJ....121.1936G} actually plotted, marked the onset
of the mass discrepancy, and pointed out discrepancies between the results of \citeauthor{2001AJ....121.1936G}  and the profiles of
\citet{2003Sci...301.1696R} for the galaxies common to both studies.

Even with the combination of stellar kinematics from  ATLAS$^\textrm{3D}$ and SLUGGS surveys we only reach radii at which dark matter just starts to be dominant -- 
and where the accelerations just decrease to values of the order of the MOND constant $a_0$. 
Using GC kinematics from the SLUGGS survey,  \citeauthor{alabi} studied dynamical models for a superset of galaxies reaching beyond $ 5\, R_{\textrm{e}}$.
The conclusions based on these should  be considered more tentative, since they are less robust than those from the JAM modelling. 
Here, we are interested in the comparison to the MOND expectations.
These were obtained by fitting the MOND profile to the JAM density profile via the (spatially constant) mass-to-light ratio as a fitting parameter.
 In Fig.~\ref{fig:profiles}, we vertically shift the density profiles of \citeauthor{alabi} so that they match the JAM profiles in the radial overlap region for comparison to the same MOND profiles. 
The \citeauthor{alabi} profiles are less smooth due to the discrete nature of the tracers, but give nonetheless some indication how the dynamically inferred density profiles in Fig.~\ref{fig:profiles} continue at larger radii. 
In some cases, e.g.~NGC~821, the deviations from MOND increase, while in other cases, e.g.~NGC~3377, the onset of the deviation appears less critical, since the continuation is close to consistent with MOND. Previous studies with dynamical tracers concluded that NGC~821 is MONDian to $\sim$3.5 $R_\textrm{e}$, using  PNe \citep{2003ApJ...599L..25M}, while \citet{2014A&A...570A.132S} could not reconcile NGC~3115 with MOND, similar to our analysis. The same applies to NGC~4278, which is in our analysis only  marginally inconsistent.

For the MDA relation, adding the  \citeauthor{alabi} data is problematic, since their profiles are non-monotonic and `jumpy', again due to the discrete nature of the tracers. Instead, we use the total masses and dark matter fractions at 5 $R_\textrm{e}$ and at $R_\textrm{max}$ from their table A4. 
 The resulting dynamical-to-stellar mass ratios and accelerations are shown in the right-hand panel of Fig.~\ref{fig:MDA}, including their 11 additional galaxies. 
The  sample of fast rotators, which overlaps with our sample of galaxies with stellar dynamics, generally appears to follow the same MDA. The even larger scatter prevents conclusions as to which interpolating function  performs better. 

However, the GC system data  
 seem also to suggest that the slow rotators have systematically higher dynamical-to-stellar mass ratios, despite the large uncertainties. 
This is qualitatively consistent with
\citet{2014A&A...570A.132S}, who generally found more severe inconsistencies with MOND for slow rotators
  based on binned kinematics of GC systems, also using SLUGGS data
(see also \citealt{2008MNRAS.387.1481A} for an account of pressure supported dwarf spheroidals within MOND).
While these tentative GC based conclusions do not share the robustness of those based on the JAM models (which were reinforced by \citealt{Serra16}), they suggest an interesting trend of increasing offsets in the MDA relation from spiral galaxies to fast rotators to slow rotators. In the  framework of MOND, these findings may be explained  by unseen (normal) matter, which  is known to be required in this context on the larger scales of galaxy groups and clusters \citep{Sanders:1999fg,2008MNRAS.387.1470A}.

In the above context it is noteworthy that \citet{2015A&A...581A..98D} found  for the ATLAS$^\textrm{3D}$ galaxies with H\,\textsc{i} data that the baryonic Tully-Fisher relation has little scatter -- which would be consistent with the MOND framework.
However,    \citet{2011ApJ...742...16T} previously found an offset  between the baryonic Tully-Fisher relation for spiral and early-type galaxies.

\section{Summary}
We analysed the mass discrepancy acceleration (MDA) relation for early-type galaxies by using the dynamical models for the stellar dynamics of 14 fast rotators of  \citet{Cappellari:2015gt}. 
The range, robustness,  and accuracy of these models allowed us to determine that,
while the galaxies broadly follow such a relation, they are systematically offset from the comparison sample of spiral galaxies \citep{2012LRR....15...10F}. This adds to the challenges found for MOND when comparing the 
dynamically determined profiles to the MOND predictions for individual galaxies.
Meanwhile, the simulations of \citet{2016MNRAS.456L.127D} demonstrated that the MDA  for spiral galaxies  could arise in  $\Lambda$CDM
 from variations of   the dark matter profile shape with galaxy mass, instead
of  a universal NFW profile. 
Our analysis predicts the corresponding models for fast-rotator early types  to be  offset from the MDA for spiral galaxies.

\section*{Acknowledgements}
We are grateful for the referee's suggestions, which improved the presentation of our results.
JJ and DAF thank the
ARC for financial support via DP130100388. MC acknowledges support from a Royal Society University Research Fellowship. 
LC was supported by the MIUR grant PRIN 2010-2011, project `The chemical and dynamical evolution of the Milky Way and Local Group Galaxies', prot. 2010LY5N2T.











\bsp	
\label{lastpage}
\end{document}